\documentclass[pra,aps,twocolumn,longbibliography,preprintnumbers,superscriptaddress,floatfix]{revtex4-2}
\usepackage[dvipsnames]{xcolor}
\usepackage[left=1.35cm,right=1.35cm,top=1.2cm,bottom=1.2cm]{geometry}
\definecolor{mediumpersianblue}{rgb}{0.0, 0.4, 0.65}
\definecolor{persianred}{rgb}{0.8, 0.2, 0.2}
\definecolor{persianorange}{rgb}{0.85, 0.56, 0.35}
\definecolor{timberwolf}{rgb}{0.86, 0.84, 0.82}
\usepackage[colorlinks=true,citecolor=persianred,linkcolor=mediumpersianblue, urlcolor=mediumpersianblue]{hyperref}
\usepackage{physics}
\usepackage{graphicx}
\usepackage{amsmath}
\usepackage{amssymb}
\usepackage{bm}
\usepackage{xspace,slashed}
\usepackage{enumitem}
\usepackage{algorithm2e}
\SetKwComment{Comment}{/* }{ */}
\usepackage{multirow}
\usepackage{booktabs}
\usepackage[utf8]{inputenc}
\usepackage{braket}
\usepackage{array}
\usepackage{tikz}
\usepackage{fontawesome}
\usetikzlibrary{quantikz}

\usepackage{fixme}

\usepackage{listings}
\usepackage{xcolor}
\definecolor{codegreen}{rgb}{0,0.6,0}
\definecolor{codegray}{rgb}{0.5,0.5,0.5}
\definecolor{codepurple}{rgb}{0.58,0,0.82}
\definecolor{backcolour}{rgb}{0.95,0.95,0.92}
\definecolor{lightgreen}{rgb}{0.56, 0.93, 0.56}
\definecolor{lightkhaki}{rgb}{0.94, 0.9, 0.55}
\definecolor{unitednationsblue}{rgb}{0.36, 0.57, 0.9}

\lstdefinestyle{mystyle}{
    backgroundcolor=\color{backcolour},
    commentstyle=\color{persianred},
    keywordstyle=\color{persianorange},
    numberstyle=\tiny\color{codegray},
    stringstyle=\color{mediumpersianblue},
    basicstyle=\ttfamily\footnotesize,
    breakatwhitespace=false,
    breaklines=true,
    captionpos=b,
    keepspaces=true,
    numbers=left,
    numbersep=5pt,
    showspaces=false,
    showstringspaces=false,
    showtabs=false,
    tabsize=2
}

\usepackage{csquotes}
\usepackage[caption=false]{subfig}
\definecolor{pyKeyword}{RGB}{106,13,173}
\definecolor{pyFunc}{RGB}{0,102,204}
\definecolor{pyString}{RGB}{34,139,34}
\definecolor{pyComment}{RGB}{120,120,120}

\lstdefinestyle{codeStyle}{
language=Python,
backgroundcolor=\color{gray!3},
basicstyle=\footnotesize\ttfamily,
keywordstyle=\color{pyKeyword}\bfseries,
stringstyle=\color{pyString},
commentstyle=\color{pyComment}\itshape,
emph={apply_gate,_apply_gate_unfold,permutations},
emphstyle=\color{pyFunc}\bfseries,
breaklines=true,
breakatwhitespace=true,
frame=single,
rulecolor=\color{gray!40},
numbers=none,
showstringspaces=false,
columns=fullflexible,
keepspaces=true,
tabsize=2,
xleftmargin=6pt,
xrightmargin=6pt,
aboveskip=6pt,
belowskip=6pt
}

\lstdefinestyle{docstringStyle}{
    language={},
    backgroundcolor=\color{gray!4},
    basicstyle=\footnotesize\ttfamily\color{black!90},
    breaklines=true,
    breakatwhitespace=true,
    frame=single,
    rulecolor=\color{gray!40},
    numbers=none,
    showstringspaces=false,
    columns=fullflexible,
    keepspaces=true,
    tabsize=2,
    xleftmargin=6pt,
    xrightmargin=6pt,
    aboveskip=6pt,
    belowskip=6pt
}

\newcommand{\Qibo}{\texttt{Qibo}\xspace}

\newcommand{\Qibolab}{\texttt{Qibolab}\xspace}
\newcommand{\Qibocal}{\texttt{Qibocal}\xspace}

\newcommand{\QiboAgent}{\texttt{QiboAgent}\xspace}

\newcommand{\NumPy}{\texttt{NumPy}\xspace}

\newcommand{\eg}{\emph{e.g.}\xspace}

\fxsetup{status=draft, layout=inline, theme=color}
\definecolor{fxnote}{rgb}{1.000,0.0000,0.0000}
\usepackage{utfsym}

\begin{document}
\title{\QiboAgent: a practitioner's guideline to open source assistants\\ for Quantum Computing code development}

\preprint{TIF-UNIMI-2026-4}

\newcommand{\MIaff}{Dipartimento di Fisica, Universit\`a degli Studi di Milano, Milan, Italy}
\newcommand{\INFN}{INFN, Sezione di Milano, I-20133 Milan, Italy}
\newcommand{\TII}{Quantum Research Center, Technology Innovation Institute, Abu Dhabi, UAE}
\newcommand{\CERNaff}{European Organization for Nuclear Research (CERN), Geneva 1211, Switzerland}
\newcommand{\ANU}{School of Computing, The Australian National University, Canberra, ACT, Australia}
\newcommand{\Boston}{Faculty of Computing \& Data Sciences, Boston University, Boston, MA, USA}
\newcommand{\Sap}{Dipartimento di Fisica, Universit\`a la Sapienza, Rome, Italy}
\newcommand{\NTU}{Division of Physics and Applied Physics, School of Physical and Mathematical Sciences, Nanyang Technological University, Singapore}
\newcommand{\HelTeq}{QTF Centre of Excellence, Department of Physics, University of Helsinki, FI-00014 Helsinki, Finland}

\author{Lorenzo Esposito}
\affiliation{\MIaff}

\author{Andrea Papaluca}
\affiliation{\MIaff}
\affiliation{\ANU}

\author{Stefano Carrazza}
\affiliation{\MIaff}
\affiliation{\INFN}
\affiliation{\TII}

\begin{abstract}
We introduce \QiboAgent, a reference implementation designed to serve as a practitioner's guideline for developing specialized coding assistants in Quantum Computing middleware. Addressing the limitations in scientific software development of general-purpose proprietary models, we explore how lightweight, open-source Large Language Models (LLMs) provided with a custom workflow architecture compare. In detail, we experiment with two complementary paradigms: a Retrieval-Augmented Generation pipeline for high-precision information retrieval, and an autonomous agentic workflow for complex software engineering tasks. We observe that this hybrid approach significantly reduces hallucination rates in code generation compared to a proprietary baseline, achieving a peak accuracy of 90.2\% with relatively small open-source models of size up to 30B parameters. Furthermore, the agentic framework exhibits advanced coding capabilities, automating the resolution of maintenance issues and new features requests, or by prototyping larger-scale refactors of the codebase, such as producing a compiled Rust module with bindings of an original pure python package, \Qibo in our case. The LLM workflows used for our analysis are integrated into a user interface and a Model Context Protocol server, providing an accessible tool for \Qibo developers. Source code and deployment instructions are publicly available at \url{https://github.com/qiboteam/qiboagent}.
\end{abstract}

\maketitle
\tableofcontents

\section{Introduction}
\label{sec:introduction}
Quantum computing relies heavily on software libraries to interface algorithms with hardware. In this work, we address the software engineering challenges associated with maintaining such ecosystems, specifically within the context of \Qibo \cite{Efthymiou_2021}, an open-source middleware for quantum computing. As the framework scales, the increasing volume of the codebase introduces significant maintenance overhead. This underscores the importance of facilitating the onboarding process for new researchers and assisting developers in the ongoing evolution of the framework.

In this context, Generative AI has emerged as a transformative tool. Large Language Models (LLMs) have demonstrated remarkable proficiency in code generation and explanation. Yet, despite their general capabilities, off-the-shelf models face inherent limitations when applied to specialized scientific software. Relying solely on parametric memory, standard models frequently suffer from hallucinations and lack the reasoning depth required for complex state-dependent engineering tasks.

To address these limitations within the quantum software domain, we present a comprehensive analysis and a practitioner's guideline for deploying domain-aware AI assistants. We ground our investigation in a detailed case study that yields \QiboAgent, which serves as a practical reference tool. Our work systematically evaluates how to integrate existing paradigms, specifically \textit{Retrieval-Augmented Generation} (RAG) and \textit{Agentic Workflows}, to achieve accuracy and autonomy. Furthermore, we prioritize the evaluation of efficient, open-source models. We show that, when augmented with a structured knowledge base and agentic tools, these lightweight architectures provide a viable solution for library maintenance and a robust reference for researchers aiming to build similar specialized assistants.

We begin by detailing the technical architecture in Sec.~\ref{sec:methods}, focusing on the implementation of the RAG pipeline and the Agentic framework. To validate this dual approach, Sec.~\ref{sec:benchmark} establishes a comprehensive evaluation protocol designed to stress-test across varying levels of complexity. We conclude in Sec.~\ref{sec:results} with a quantitative and qualitative analysis of the system's performance on these maintenance and development tasks.

\section{Related Work}

The application of Large Language Models to scientific computing and physics has recently gained significant traction. While general-purpose models demonstrate strong zero-shot capabilities, their performance in highly specialized domains often remains limited without domain adaptation or external knowledge grounding.

In astrophysics, domain-adapted models such as AstroLLama \cite{nguyen2023astrollamaspecializedfoundationmodels} and AstroLLamaChat \cite{perkowski2024astrollamachatscalingastrollamaconversational} have been fine-tuned on curated literature corpora, enabling more accurate responses to astrophysics-specific queries. Similarly, in High Energy Physics, systems like MadAgents \cite{plehn2026madagents} adopt a multi-agent architecture based on proprietary models to guide users through the MadGraph workflow, from installation to event generation.

In the context of quantum computing, Dupuis et al.~\cite{dupuis2024qiskitcodeassistanttraining} extend the pre-training of LLMs using the Qiskit codebase to construct a specialized quantum coding assistant, demonstrating improved performance over general-purpose models in custom benchmarks. This line of work highlights the effectiveness of domain-specific pre-training but relies on heavyweight training pipelines and static parametric memory.

Retrieval-Augmented Generation (RAG) has been explored as a strategy to improve the factual grounding of LLMs in quantum programming. PennyLang \cite{basit2025pennylangpioneeringllmbasedquantum} introduces both a curated quantum code dataset and a RAG-based pipeline that significantly improves code generation accuracy over non-augmented baselines. However, their focus is primarily on language-level code synthesis rather than on real-world software maintenance or developer workflows.

More recently, autonomous agentic workflows are emerging to tackle complex design problems within the quantum domain. For instance, recent work \cite{Knipfer_2026kng} introduces an agent-based framework for optimizing Variational Quantum Circuit (VQC) architectures, demonstrating how agents can iteratively propose, evaluate, and refine quantum models with minimal human intervention.

\QiboAgent is designed as a reference architecture for deploying domain-aware coding assistants that integrate two complementary paradigms: structured knowledge grounding through Retrieval-Augmented Generation and autonomous problem solving through agentic workflows. Rather than relying on domain-specific fine-tuning, our approach leverages lightweight open-source models coupled with a knowledge base, avoiding costly retraining procedures. This design enables \QiboAgent to operate both as an interactive coding assistant for end users and as a development tool for maintainers, supporting tasks such as documentation synthesis, automated pull request resolution, and codebase refactoring.

\section{Different approaches to LLMs}
\label{sec:methods}

In recent years, the landscape of Artificial Intelligence has been fundamentally reshaped by the proliferation of Large Language Models, a shift catalyzed by the introduction of the Transformer architecture \cite{vaswani2023attentionneed}. These models have exhibited substantial cross-domain adaptability, specifically becoming integral collaborative assistants within modern programming workflows.

At their core, standard LLMs operate on a relatively simple mechanism: probabilistic token inference. They generate responses by predicting the most likely subsequent token based on an input prompt. However, while this mechanism is powerful, relying solely on a model's internal weights has inherent limitations regarding reasoning depth and factual accuracy \cite{huang2025survey}. To address these constraints, the research community has developed various methodological techniques designed to augment the core model's capabilities. Common examples include \textit{Chain-of-Thought (CoT) prompting} \cite{wei2023chainofthoughtpromptingelicitsreasoning}, which encourages the model to decompose complex problems into intermediate reasoning steps, and \textit{Few-Shot Learning} \cite{brown2020languagemodelsfewshotlearners}, which provides context-specific examples within the prompt to guide generation.

In the development of \QiboAgent, we integrated two distinct methodologies to refine the model into a robust coding assistant:
\begin{itemize}
    \item \textbf{Retrieval-Augmented Generation:} Adopted to overcome the limitations of static parametric memory. By connecting the model to a Knowledge Base, this approach mitigates hallucinations and ensures that code generation is conditioned on the actual \Qibo codebase.
    \item \textbf{Agentic Workflows:} Employed to evolve the LLM from a text generator into an autonomous controller. Equipped with executable tools, the model can actively interact with the codebase to perform multi-step engineering tasks, such as collecting information and iteratively fixing execution errors.
\end{itemize}

By leveraging these two paradigms, the underlying generative models transition from a general-purpose coder to a domain-aware collaborator. We first detail the RAG infrastructure designed for knowledge management (Sec.~\ref{sec:RAG}), followed by the agentic framework developed for complex software engineering tasks (Sec.~\ref{sec:Agentic}).

\subsection{System Goals}
\label{sec:Goals}

The primary objective of \QiboAgent is to provide a versatile, multi-role assistant capable of operating across different layers of the software lifecycle. Rather than serving a single use case, the integration of RAG and agentic workflows enables the system to adapt its behavior to the specific needs of different users within the \Qibo ecosystem. We conceptualize these operational modalities as follows (summarized in Fig.~\ref{fig:modalities}):

\begin{figure}[htbp]
    \centering
    \includegraphics[width=\columnwidth]{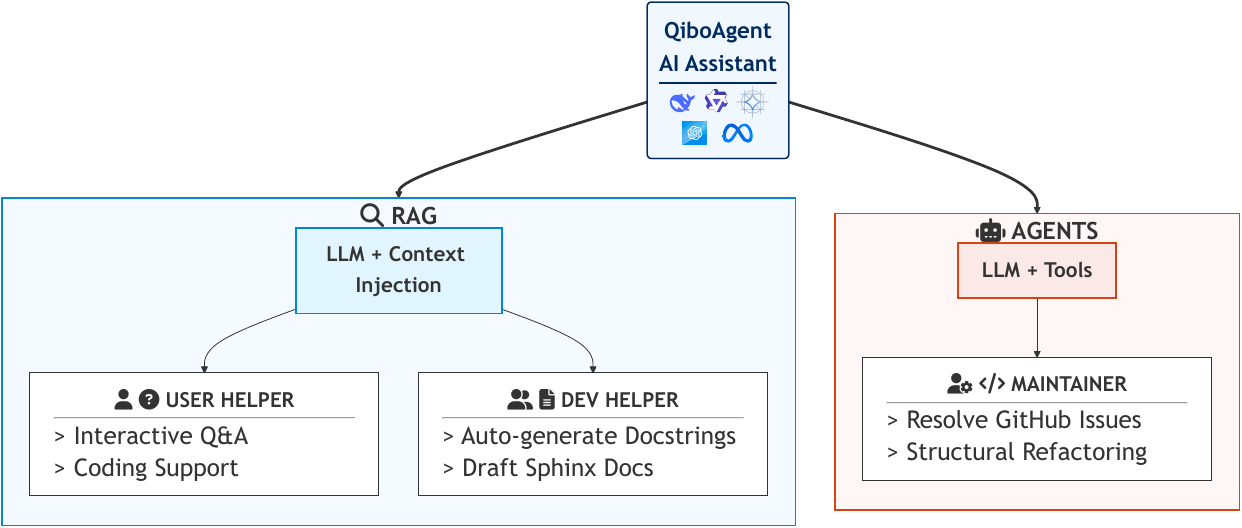}
    \caption{\textbf{Multi-role operational modalities of \QiboAgent}. The framework encompasses two distinct operational layers: (Left) a RAG-based assistance layer providing context-aware support for users and developers (\eg, interactive Q\&A and automated documentation); (Right) an autonomous Agentic layer designed for repository maintenance, enabling proactive GitHub issue resolution and structural code refactoring.}
    \label{fig:modalities}
\end{figure}

\begin{itemize}
    \item \textbf{User Helper:} Designed to flatten the learning curve for new users. Utilizing the RAG pipeline, the system acts as an interactive Q\&A chatbot whose responses are grounded in the codebase to support code development.
    \item \textbf{Developer Helper:} Aimed at streamlining routine coding tasks. This modality utilizes contextual retrieval to generate syntactically accurate docstrings, enforce stylistic conventions, and draft structural documentation for new modules.
    \item \textbf{Maintainer:} Targeted at complex repository management. Through the agentic framework, the system autonomously addresses GitHub issues and suggests code patches.
\end{itemize}

These distinct modalities dictate the evaluation benchmarks presented in Section \ref{sec:benchmark}, ensuring that both the RAG and agentic components are tested against realistic operational scenarios.

\subsection{Retrieval Augmented Generation}
\label{sec:RAG}

Despite their proficiency in code and natural language generation, LLMs exhibit an inherent susceptibility to hallucinations. To address this problem in developing \QiboAgent, we employed Retrieval-Augmented Generation \cite{lewis2021retrievalaugmentedgenerationknowledgeintensivenlp} (RAG), a technique designed to enhance model reliability by grounding its outputs in verifiable, external information.

The process begins with the construction of a Knowledge Base, a comprehensive repository containing domain-specific structured data, such as technical documentation, source code, or internal records, that encompasses all the information relevant to a specific field of application.

Once this repository is assembled, the pipeline begins with an ingestion phase, where these documents are partitioned into discrete segments, or \textit{chunks}, to align with the model's finite context window. To prevent semantic truncation across boundaries, this segmentation typically employs a sliding window overlap, ensuring the preservation of contextual continuity. These text segments are subsequently transformed by an embedding model into dense vectors within a high-dimensional latent space, where geometric proximity encodes semantic similarity.

During the inference phase, the user's query undergoes an identical vectorization process. A retrieval algorithm then performs a similarity search (\eg, via cosine similarity) over the vector-store to identify the top-$k$ segments most relevant to the query. Finally, this retrieved context is injected directly into the LLM's prompt, effectively conditioning the generation on retrieved evidence and significantly mitigating the risk of hallucinations.

\begin{figure*}[ht]
    \centering
    \includegraphics[width=2.0\columnwidth]{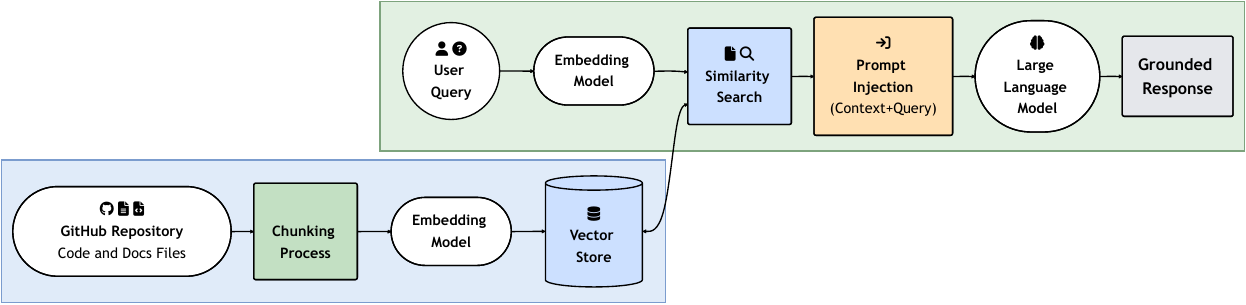}
    \caption{\textbf{RAG Pipeline Architecture.} The process begins with the construction of a knowledge base filtering code files and documentation from the Github repository. During the ingestion phase, documents are segmented into chunks and embedded into a vector space. At inference time, the user's query is vectorized and used to retrieve relevant context, which is then injected into the LLM's prompt to condition generation.}
    \label{rag_diagram}
\end{figure*}

For the development of \QiboAgent, we evaluated distinct RAG approaches built upon a comprehensive knowledge base scraped directly from the official Qibo GitHub repository. This dataset encompassed all relevant source code files (including Python scripts and Jupyter notebooks) as well as full documentation assets (Markdown and reStructuredText).

\subsubsection{Semantic Retrieval}
\label{sec:semantic_retr}
In our first experimental configuration, we adopted a fixed-size chunking strategy, where specific chunk lengths and overlap values were defined according to the file type (\eg, differing parameters for Python scripts versus Markdown or reStructuredText documentation).
To refine the retrieval quality, we enhanced the standard vector search with a two-stage reranking mechanism designed to minimize informational redundancy. Standard vector retrieval often suffers from "semantic clustering", where the top retrieved results are slight variations of the same information rather than diverse aspects of the query. To mitigate this, our pipeline first retrieves a broad candidate set of $N=24$ vectors based strictly on cosine similarity. Subsequently, we apply the Maximum Marginal Relevance \cite{MMR} (MMR) algorithm to select the final top-$k$ context chunks. MMR re-ranks each candidate by maximizing its similarity with the input query while minimizing the similarity to the other documents already selected. Formally, the next document $d^*$ to be added to the context set $S$ is selected according to:

\begin{multline}
{\rm MMR} \coloneqq \operatorname*{argmax}_{d_i \in C \setminus S} \Big[ \lambda \cdot \text{Sim}(d_i, q) \\
- (1 - \lambda) \cdot \max_{d_j \in S} \text{Sim}(d_i, d_j) \Big],
\end{multline}
where $C$ is the set of candidate documents, $q$ is the query vector, and $\text{Sim}$ denotes the vector similarity. The hyperparameter $\lambda \in [0, 1]$ controls the diversity trade-off: a value of $\lambda = 1$ yields standard relevance ranking, while lower values penalize redundancy. In our experiments, we tuned this parameter to $\lambda = 0.5$ ensuring the LLM receives a context that is both pertinent and informationally diverse.

\subsubsection{Hybrid Retrieval}
\label{sec:hybrid_retr}
While the semantic approach captures the conceptual intent of a query, it often struggles with the high-precision requirements of a specialized quantum computing library like \Qibo. The framework heavily relies on domain-specific terminology and scientific nomenclature, such as specific Hamiltonian identifiers (\eg, Transverse-Field Ising Model, Heisenberg $XXX$) or variational ansatz definitions. Dense vector embeddings, optimized for general semantic proximity, may occasionally overlook these exact technical tokens or conflate distinct physical models. To overcome these limitations and ensure the retrieval of exact matches, we implemented a second, more robust configuration.

This approach was supported by a refined structure-aware ingestion pipeline. Deviating from the brute-force sliding window method, we developed a custom code splitter designed to respect the syntactic boundaries of the source code. Instead of arbitrarily splitting text, the system constructs hierarchically meaningful chunks: class definitions are aggregated with their signatures and constructors, while individual methods are isolated as distinct units enriched with metadata. This ensures that even when isolated, code snippets retain strong semantic links to their parent structures.

To leverage this structured data, the retrieval logic synergizes the conceptual understanding of dense vector embeddings with the lexical precision of keyword matching via the Okapi BM25 algorithm \cite{BM25}. We implemented a rank-aware normalized fusion mechanism to integrate the bounded vector similarity scores with the unbounded term-frequency metrics of BM25. Since raw similarity scores and BM25 scores operate on incompatible scales, we mapped both to a normalized range $[0,1]$ before aggregation. Specifically, the normalized lexical score is obtained by scaling the raw BM25 scores relative to the maximum score in the batch. Conversely, for the semantic component, we utilized a linear rank-based metric rather than raw cosine similarity to prioritize the relative ordering of the entire document collection. The final hybrid relevance score is derived via a weighted linear combination:

\begin{equation}
S_{\text{hybrid}} = \alpha \cdot \frac{S_{\text{BM25}}}{\max(\mathbf{S}_{\text{BM25}})} + (1 - \alpha) \cdot \left(1 - \frac{r}{N}\right),
\end{equation}
where $S_{\text{BM25}}$ is the raw keyword score, $\max(\mathbf{S}_{\text{BM25}})$ is the maximum BM25 score in the retrieved batch, $r$ is the zero-indexed rank of the document in the vector search results, and $N$ is the total number of documents in the knowledge base. The hyperparameter $\alpha \in [0,1]$ balances the trade-off between strict keyword matching and semantic context. To maintain a neutral stance between strict keyword matching and semantic retrieval, we considered $\alpha = 0.5$ to be a reasonable baseline. This balanced default proved effective for our specific knowledge base.

\subsection{Agentic approach}
\label{sec:Agentic}
While retrieval mechanisms successfully address the need for static knowledge, standard inference remains insufficient for complex engineering tasks requiring state management and multi-step reasoning. To bridge this gap, we implemented an Agentic Workflow, a paradigm that transforms the LLM from a passive text generator into an active controller capable of dynamic interaction with the codebase. The model operates within a cognitive architecture based on the ReAct framework \cite{yao2023reactsynergizingreasoningacting}, generating interleaved reasoning traces and action-specific tokens. This allows the system to decompose high-level objectives into executable steps, effectively mimicking the iterative problem-solving workflow of a human developer.

We customized this environment to address two distinct operational scenarios. For repository maintenance, we deployed a single-agent architecture equipped with a suite of investigative tools, including a GitHub issue scraper and a file system navigation module. This setup enables the agent to autonomously reconstruct the problem context: it parses conversation histories to understand reported bugs and traverses the directory structure to locate and read relevant files. Finally, the agent outputs a comprehensive solution, providing both a brief explanation of the fix and the specific code patch needed to resolve the issue.

Conversely, the modernization of \Qibo with its core module demanded a Sequential Multi-Agent Architecture (Fig.~\ref{fig:core_diagram}). To mitigate context window saturation inherent in generating extensive codebases, we decomposed the workflow into discrete stages, instantiating distinct agents to sequentially synthesize the compiled Rust backend, the corresponding Python bindings, and the verification suite. These agents operate within a robust self-correcting feedback loop. Unlike standard generation where errors are terminal, this system is empowered with shell execution capabilities: agents can trigger compilation commands (\eg, \texttt{cargo build}), capture standard error logs upon failure, and use this feedback to diagnose and iteratively refine the code. This cycle, spanning drafting, compilation, observation, and correction, enables the resolution of complex logical errors that a single inference pass cannot guarantee.

\subsection{Model selection}
The operational backbone of this reference implementation is based on open-source LLMs. We use the Ollama framework as the local inference engine to facilitate seamless deployment, while the orchestration of the Retrieval-Augmented Generation pipelines and agentic workflows is implemented using LangChain.

To power the semantic retrieval mechanism, we utilize the \texttt{sentence-transformers/all-MiniLM-L6-v2} model \cite{reimers2019sentencebertsentenceembeddingsusing} as the embedding engine, projecting textual data into a 384-dimensional dense vector space. This model was selected as it represents a standard, lightweight choice that proved highly effective during our initial RAG configuration tests. It provides semantic accuracy for our use case while ensuring low-latency retrieval without the computational overhead of larger embedding models, aligning perfectly with our goal of a fully local infrastructure.

Regarding the generative component, the setup supports a tiered selection strategy, matching model capacity to task complexity:

\begin{itemize}
    \item \textbf{Efficiency-Oriented Lightweight Models:} For standard operations, including \Qibo code-related queries and technical documentation generation, we utilized a range of lightweight models with scales up to 70B parameters, such as \texttt{gpt-oss:20b} and \texttt{qwen3-coder:30b}. Through our RAG-based enhancements, these models demonstrated high accuracy and grounded reasoning while maintaining a significantly lower computational footprint compared to larger architectures.

    \item \textbf{Large-Scale Agentic Reasoning:} For high-complexity tasks, which demand superior reasoning and multi-step planning, we selected larger models such as \texttt{gpt-oss:120b}. Although these are significantly more compact than proprietary models (\eg, GPT-4 from OpenAI or Gemini 2.5 from Google, which are estimated to reach the trillion-parameter scale), they were chosen for their robust native support for tool-calling, a prerequisite for the LLM to act as the central controller of the agentic system.
\end{itemize}

To ensure experimental reproducibility and maximize result determinism, the temperature hyperparameter was fixed at $T=0$ across all evaluations. Detailed specifications for the models used in this study, including their parameter counts and specialized domains, are summarized in Table~\ref{table:models}.

\begin{table*}[t]
\centering
\begin{tabular}{lccccc}
\toprule
\textbf{Model Name} & \textbf{Parameters} & \textbf{Type} & \textbf{Task Allocation} & \textbf{Quantization} & \textbf{VRAM} (GB) \\
\midrule
\texttt{all-MiniLM-L6-v2}~\cite{reimers2019sentencebertsentenceembeddingsusing} & 22M & Embedding & RAG Embeddings & -- & -- \\
\texttt{gpt-oss}~\cite{openai2025gptoss120bgptoss20bmodel} & 20B & Generative & Q\&A & MXFP4 & $\sim$ 17 \\
\texttt{gemma3}~\cite{gemmateam2025gemma3technicalreport} & 27B & Generative & Q\&A & Q4\_K\_M & $\sim$ 20 \\
\texttt{qwen3-coder}~\cite{qwen3technicalreport} & 30B & Generative & Q\&A, Documentation, Agentic Workflows & Q4\_K\_M &  $\sim$ 22 \\
\texttt{deepseek-coder}~\cite{guo2024deepseekcoderlargelanguagemodel} & 33B & Generative & Q\&A, Documentation & Q4\_0 & $\sim$ 22 \\
\texttt{deepseek-r1}~\cite{Deepseek_2025} & 70B & Generative & Q\&A, Documentation & Q4\_K\_M & $\sim$ 48 \\
\texttt{gpt-oss}~\cite{openai2025gptoss120bgptoss20bmodel} & 120B & Generative & Agentic Workflows & MXFP4 & $\sim$ 72 \\
\bottomrule
\end{tabular}
\caption{Specifications of the models employed in \QiboAgent. VRAM was estimated during inference phase, it may vary based on the context length and the specific task.}
\label{table:models}
\end{table*}

\section{The benchmark tasks}
\label{sec:benchmark}

To assess the capabilities of the \QiboAgent framework as a comprehensive assistant for quantum computing code, we designed a multi-faceted evaluation campaign. These benchmarks target two primary stakeholders: end-users seeking guidance on code implementation, and developers tasked with library maintenance and modernization.
Accordingly, we defined four specific benchmark tasks, ranging from standard information retrieval to complex autonomous software engineering, to test the system's limits across different levels of abstraction.

\subsection{Quantum Computing Q\&A}
\Qibo is an open-source full-stack API for quantum simulation and quantum hardware control. It is fundamentally built upon two guiding principles: simplicity, through an agnostic design to quantum primitives, and flexibility, via a transparent mechanism to execute code across both classical and quantum hardware.
Building upon this foundation of accessibility, \QiboAgent is introduced as an additional support layer designed to assist new users. While the framework itself simplifies the technical interaction with quantum computing, \QiboAgent aims to streamline the learning curve of the coding process.

The capabilities of \QiboAgent in assisting new users were evaluated through an experimental pipeline assessing three distinct retrieval configurations: a No-RAG Baseline (zero-shot inference relying on parametric memory), the Semantic Retrieval (\ref{sec:semantic_retr}) strategy and the Hybrid Retrieval (\ref{sec:hybrid_retr}) approach (combining vector similarity with keyword matching).

Experimental consistency was maintained by subjecting every model to an identical, standardized system prompt for every evaluated question, ensuring a fair comparison of the retrieval logic. The full list of prompts used in these benchmarks is available in the project repository. Furthermore, the pipeline's performance was benchmarked against the commercial \texttt{Gemini-2.5-flash} platform, utilizing its native retrieval capabilities to establish a competitive baseline.

\subsubsection{Query Dataset}

We established a curated dataset \footnote{Dataset available at \url{https://github.com/qiboteam/qiboagent}} comprising 50 questions designed to ensure comprehensive coverage of the \Qibo ecosystem. The samples span across diverse modules, ranging from:
\begin{itemize}
    \item \textbf{Configuration:} Queries regarding backend selection (\eg, switching to \NumPy backend), precision settings, and thread management.
    \item \textbf{Core Primitives Logic:} Queries ranging from basic circuit construction to complex flow control. This includes creating specific quantum states (\eg, Bell, GHZ), managing measurements, and visualizing circuits.
    \item \textbf{Advanced Simulation:} Queries testing the model's ability to handle density matrices and open quantum systems. Questions explicitly require implementing \textit{Kraus channels}, \textit{depolarizing channels}, and extracting entanglement entropy.
    \item \textbf{Algorithmic Implementation:} High-level queries requiring the use of full quantum algorithms, including VQE, QAOA, and Grover's search. These queries test the integration of Hamiltonians (\eg, TFIM, Heisenberg) with circuit execution.
    \item \textbf{Debugging and Software Engineering:} Queries range from debugging broken code snippets to generating \textit{unit tests} and custom classes inheriting from \Qibo modules.
\end{itemize}

To validate the model outputs, we generated a comprehensive ground truth dataset by pre-calculating the expected execution results for each query, such as the final state vectors and expectation values of the quantum circuits. The actual outputs generated by the executed code were then systematically compared against this baseline, ensuring that the evaluation reflects functional correctness rather than textual similarity.

\subsubsection{Evaluation Metrics}

We evaluated the generated code snippets by monitoring three distinct performance indicators:
\begin{enumerate}
    \item \textbf{Functional Correctness:} Rather than performing a simple textual comparison, we executed the generated code in a controlled environment and cross-referenced its numerical output against the pre-calculated reference values. This methodology ensures that the reported accuracy reflects the actual functional correctness of the code rather than its mere syntactic plausibility.
    \item \textbf{Hallucination Score:} We introduced a custom metric to specifically quantify model reliability. Under this metric, runtime errors such as \texttt{AttributeError}, \texttt{ImportError}, and \texttt{ModuleNotFoundError} were flagged as evidence of hallucination. This allowed us to distinguish between \textit{logical errors}, such as applying a gate to the wrong qubits, and \textit{structural hallucinations}, where the model fabricates non-existent \texttt{qibo} modules, functions, or parameters. By tracking whether an execution failure was caused by these specific exceptions, we were able to assign a numerical value to the model's tendency to invent technical artifacts, providing a measure of its factual reliability within the domain.
    \item \textbf{Stylistic Quality:} We recorded Pylint scores to assess the stylistic quality and readability of the generated scripts.
\end{enumerate}

These metrics ensure that the models are assessed not only on their ability to produce functionally accurate code, but also on their factual reliability and adherence to established software engineering standards.

\subsection{Automated Documentation}
High-quality documentation is a primary determinant of a framework's adoption and longevity. To facilitate this, \Qibo relies on Sphinx \cite{sphinx_web}, an engine designed to compile documentation directly from source code docstrings using reStructuredText (RST) directives. Although Sphinx efficiently handles metadata extraction and building, the burden of content creation persists. 
This presents a demanding challenge: writing granular, technically accurate function-level docstrings that correctly incorporate the precise RST directives and cross-referencing logic required to build the documentation. Even when developers employ general-purpose conversational AI to expedite drafting, such generic systems remain context-agnostic. Consequently, they fail to replicate the specific formatting conventions and internal links required by a specialized repository like \Qibo. To maintain structural and stylistic continuity, a context-aware approach is essential.

Our approach addresses this by evaluating the \QiboAgent RAG pipeline within an automated workflow specifically designed for generating documentation for newly introduced code. Unlike standard solutions, this approach leverages the Hybrid RAG pipeline described in Section \ref{sec:RAG} to inject context from the existing codebase. This retrieval mechanism is critical for minimizing hallucinations in internal cross-references: by grounding the generation in actual repository data, the model can accurately link arguments to specific objects (\eg, \texttt{qibo.models.circuit.Circuit}). This approach also guarantees that synthesized usage examples are syntactically consistent with the current API and compliant with Sphinx formatting standards.

Our empirical validation of the system's capabilities focused on function-level docstring generation. Using our RAG-based approach, we generated descriptions for multiple functions and classes across various existing modules. We verified that the model successfully utilized the retrieved context to produce coherent documentation and Sphinx-compatible formatting, effectively mimicking the stylistic conventions of the existing library.
During this analysis, we enforced a strict data decontamination protocol to guarantee experimental integrity. The source file containing the target functions was temporarily removed from the RAG knowledge base. This step prevented the model from simply retrieving and copying existing docstrings, ensuring that the generated output was a genuine synthesis based purely on the underlying code logic and the architectural context retrieved from the rest of the repository.

In addition to function-level descriptions, the same pipeline can be naturally extended to synthesize structural Sphinx documentation. As a practical demonstration, we employed the system to draft the \texttt{.rst} files for the \texttt{qibo.ui} module, which was previously missing from the official \Qibo API reference.

\subsection{New Feature Pull Request}\label{sec:newfeaturepr}

\begin{figure}[h!]
    \centering
    {\includegraphics[scale=0.45]{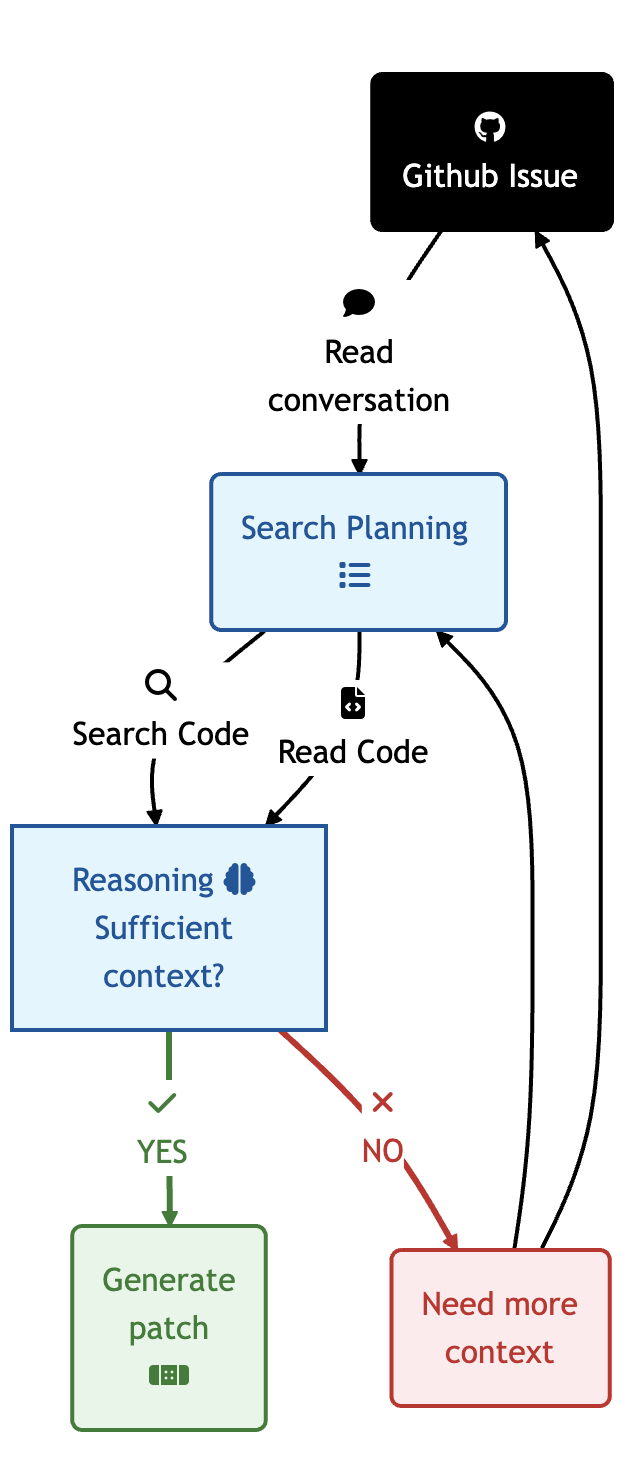}}
    \caption{Schematic representation of the agentic workflow for automated issue resolution. The diagram illustrates the iterative ReAct loop, where the agent utilizes a specialized toolset to analyze issue specifications, navigate the repository structure, and synthesize code patches.}
    \label{fig:diagram}
\end{figure}

Moving beyond user support and documentation synthesis, we evaluated the \QiboAgent architecture's capacity to actively participate in the repository's maintenance lifecycle. Leveraging the agentic framework described in Section \ref{sec:Agentic}, we designed an experiment to simulate the development process of a human contributor addressing official GitHub issues. The agent was equipped with a specialized toolset allowing it to scrape conversation histories from the issue tracker and autonomously navigate the codebase structure. Through a "Reason-Act" iterative loop, visualized in Figure \ref{fig:diagram}, the system analyzes the requirements, identifies the relevant files, and formulates a plan before attempting any code modification.

To empirically validate this workflow across different levels of technical difficulty, we submitted two distinct tasks from the official \Qibo repository to the agent. We first introduced a lower-complexity task (Issue \#1699), which required updating the high-level \texttt{Circuit.set\_parameters} method to support non-trainable parameters, effectively testing the agent's ability to handle \Qibo logic. The system successfully resolved this issue by autonomously localizing the correct source file and implementing the necessary filtering logic without human intervention. Subsequently, we challenged the system with a more articulate task (Issue \#1710) involving the \texttt{apply\_gate} function within the \NumPy backend. This second scenario required a deep understanding of quantum state manipulation and backend architecture, serving as a stress test for the agent's ability to gain context and synthesize a patch that is not only syntactically correct but also consistent with the library's physics simulation standards. To further assess the agent's autonomy in this complex scenario, we tested it against two variations of the issue description: since the original thread contained a code template for the proposed algorithm, we conducted two distinct execution runs. In the first run, the agent autonomously processed the complete original conversation; in the second, it processed a redacted version where the guiding snippet was removed.

\subsection{Refactoring and Rebasing}

The final and most complex benchmark addresses a critical challenge in the scientific software lifecycle: the modernization of legacy codebases towards high-performance compiled languages. Specifically, we tasked the \QiboAgent multi-agent system with the "greenfield" development of \texttt{Qibo-core}, a core module written in Rust. This component is designed as a language-agnostic kernel serving as a foundation for generating bindings across multiple high-level languages, starting with the Python interface.

Unlike previous maintenance tasks limited to single-file editing, this challenge necessitates the generation of a comprehensive, multi-file project structure from scratch. The primary objective was to synthesize a functional Proof of Concept (PoC) that, while not targeting immediate feature parity with the full library, establishes a robust architectural foundation. In particular, the system was required to:
\begin{enumerate}
    \item Implement fundamental quantum primitives (Gates and Circuits) in Rust;
    \item Generate the essential build configuration files (\eg, \texttt{Cargo.toml}, \texttt{pyproject.toml}) alongside a suite of unit tests;
    \item Create Python bindings using the \texttt{pyO3} library and a high-level Python wrapper that strictly preserves the syntax of the existing \Qibo API.
    \item Execute the generated \texttt{Circuit} object through the \Qibo \NumPy backend to demonstrate the seamless compatibility of the Rust core and its Python bindings with the original API.
\end{enumerate}

To address the inherent context window limitations associated with generating an entire library ecosystem, we transitioned from single-turn inference to a Sequential Multi-Agent Architecture. In this framework, the generative process is decomposed into specialized roles, each responsible for a distinct layer of the software stack:

\begin{figure*}[htbp]
    \centering
    {\includegraphics[width=\textwidth]{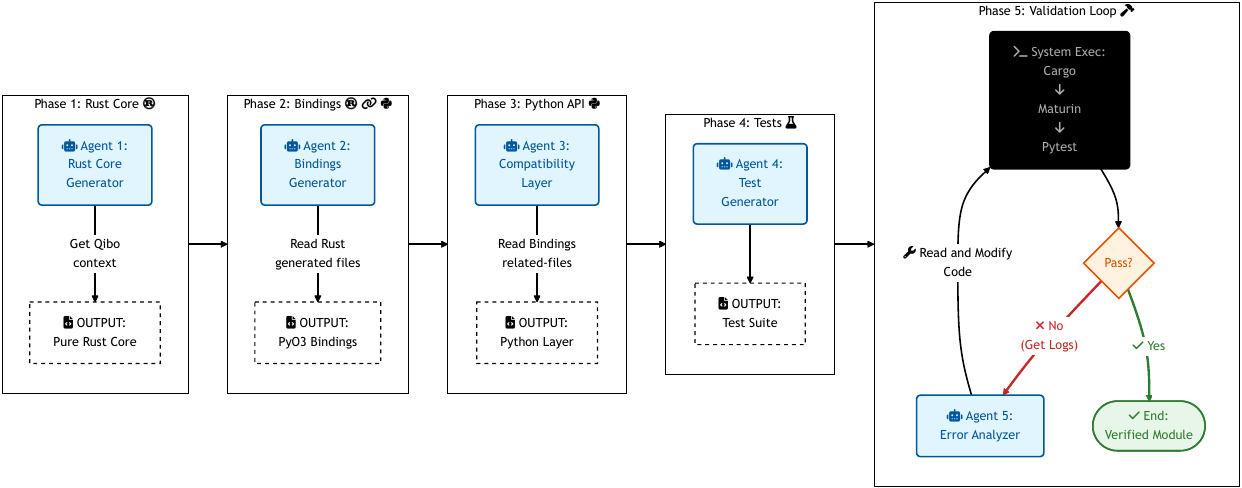}}
    \caption{Schematic representation of the multi-agent architecture for the development of the \texttt{Qibo-core} module. The diagram illustrates the sequential workflow, where each agent is responsible for a specific layer of the software stack, from Rust implementation to Python bindings and validation.}
    \label{fig:core_diagram}
\end{figure*}

\begin{itemize}
    \item \textbf{Rust Architect Agent:} Responsible for implementing the low-level logic, defining structs for quantum gates and circuits in pure Rust.
    \item \textbf{Bindings Agent:} Tasked with developing the binding layer, translating Rust types into Python-compatible objects via \texttt{pyO3}, and managing the build system setup.
    \item \textbf{Compatibility Agent:} Generates the high-level Python compatibility layer, ensuring that the new module's syntax remains identical to the original package.
    \item \textbf{Validation Agent:} Dedicated to synthesizing comprehensive unit tests that validate the functional equivalence and API compatibility between the newly compiled core module and the original \Qibo framework.
\end{itemize}
Finally, to ensure functional correctness, we integrated a Compiler-Driven Feedback Loop. A dedicated \textbf{Debugger Agent} is granted the capability to visualize shell command outputs(\eg, \texttt{cargo build}, \texttt{pytest}). Upon encountering compilation errors or test faults, this agent parses the standard error output, retrieves the relevant source files, and autonomously applies corrective patches to the codebase.

\section{Results}
\label{sec:results}

\subsection{Questions benchmark}
\begin{figure*}[htbp]
    \centering
    \includegraphics[scale=0.4]{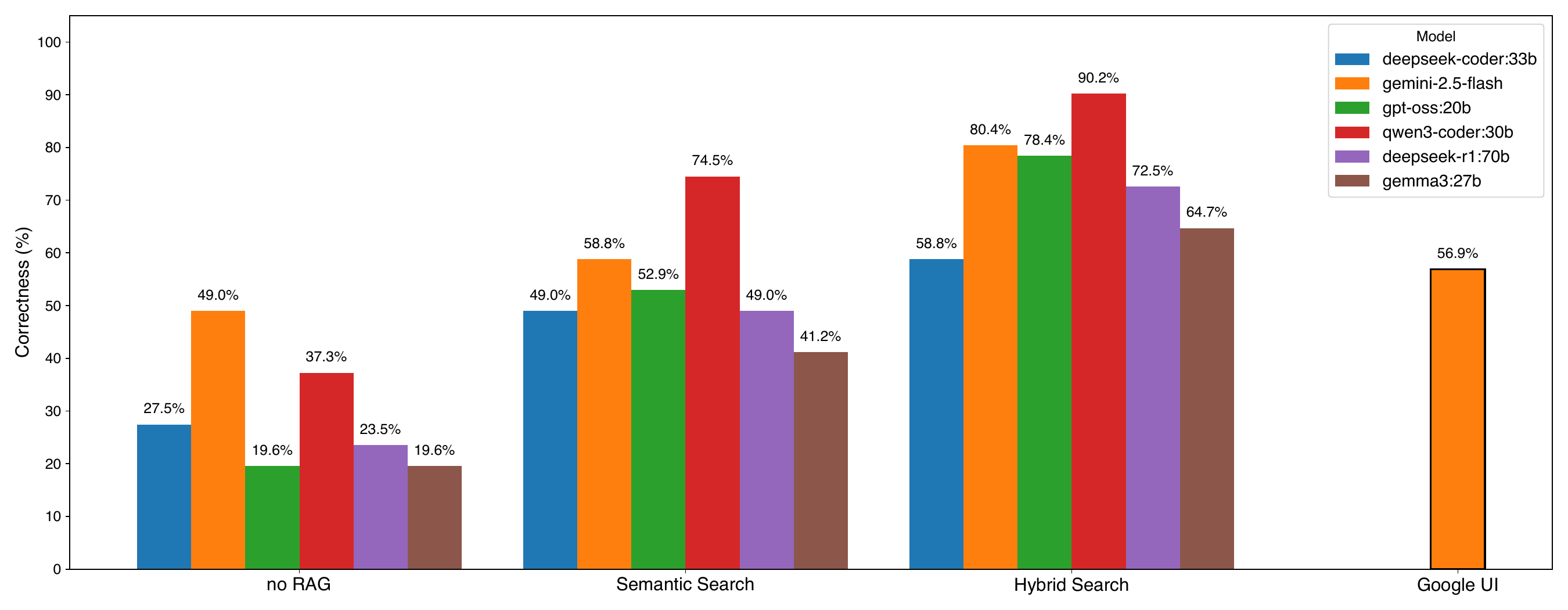}
    \caption{\textbf{Code generation accuracy on the 50-question benchmark.} We compare six LLMs across three retrieval configurations (No-RAG, Semantic Search, Hybrid Search) against the commercial Google Native UI baseline ($56.9\%$). The Hybrid Retrieval strategy consistently yields the highest performance, peaking at $90.2\%$ with \texttt{qwen3-coder:30b}. Notably, this setup allows lightweight open-source models ($20\text{--}30$B parameters) to significantly outperform the general-purpose commercial baseline, demonstrating the dominance of domain-specific context over raw model scale.}
    \label{fig:diff_approach}
\end{figure*}

As reported in Figure \ref{fig:diff_approach}, the results demonstrate a clear performance hierarchy. The No-RAG baseline serves as a lower bound, confirming that models struggle to generate correct \Qibo code without external context. While Semantic Search provides substantial improvements, the Hybrid Retrieval strategy maximizes performance across all models.

The standard Google Native web interface benchmark yields an accuracy of $57\%$, roughly equivalent to our intermediate Semantic Search performance. Interestingly, a closer inspection reveals a notable anomaly regarding \texttt{Gemini-2.5-flash}: its No-RAG baseline significantly outperforms the other models, yet it exhibits minimal improvement when transitioning to the Semantic Search pipeline. This behavior strongly suggests that the model has likely encountered the \Qibo codebase during its pre-training phase, relying on its internal parametric memory rather than retrieved context for basic queries. Despite this initial pre-training advantage, applying our Hybrid approach to the same \texttt{Gemini-2.5-flash} model boosts accuracy to $80\%$. This gap highlights that a specialized retrieval strategy remains more critical than raw model size or pre-training exposure for domain-specific code generation.

\begin{figure*}[htbp]
  \centering
  \fbox{\includegraphics[width=\textwidth]{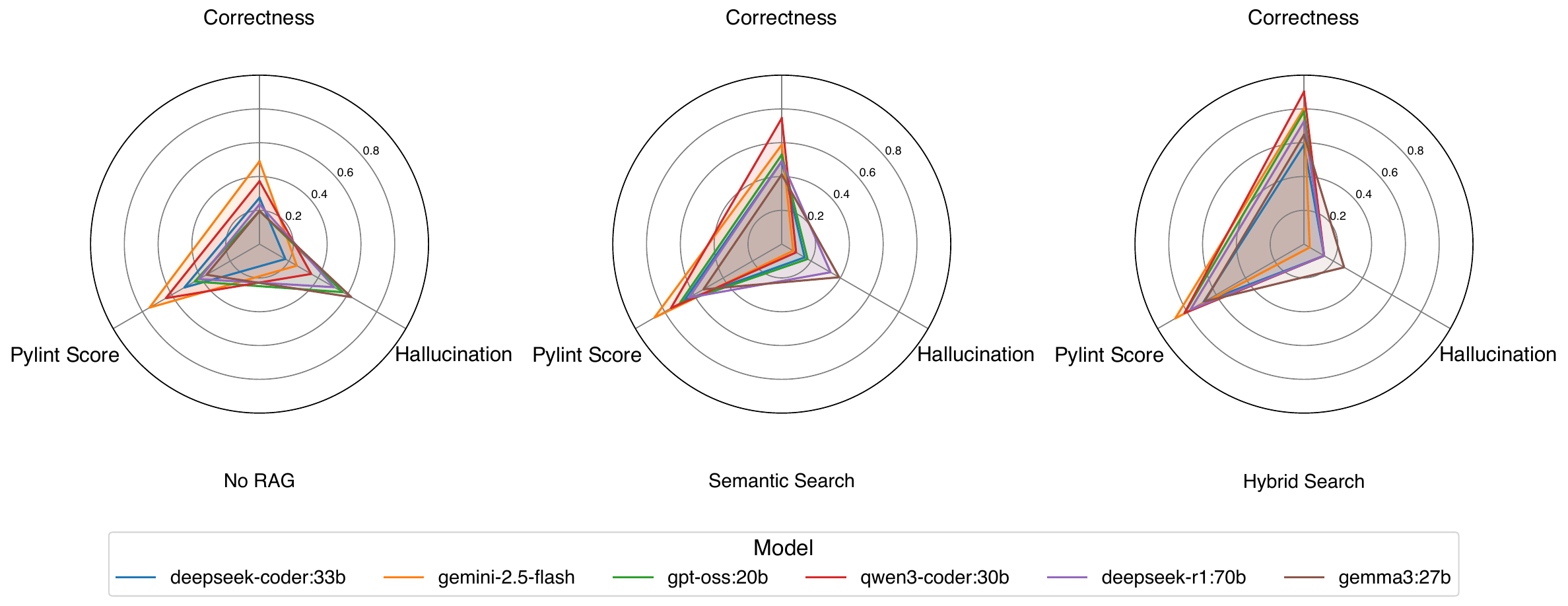}}
\caption{\textbf{Impact of RAG configurations on model generation quality.} The radar plots compare the baseline (No RAG) against Semantic and Hybrid Search retrieval methods across different models, evaluated on a dataset of 50 questions. Axes represent Correctness, Pylint Score and Hallucination (where 1 indicates a detected hallucination). The transition from baseline to hybrid-RAG-based approaches yields a significant reduction in hallucination rates. Furthermore, the marginal improvement in Pylint scores indicates that providing context via retrieval not only grounds the logical reasoning but also acts as a syntactic template, aiding the models in adhering to the \Qibo library's native coding style.}
\label{fig:radar_plot}
\end{figure*}

Figure \ref{fig:radar_plot} further demonstrates that the RAG-based approach yields a marked decrease in hallucination rates compared to the baseline. Concurrently, we note a marginal increase in the Pylint quality score. This suggests that the presence of retrieved documents not only grounds the logic but also serves as a syntactic template, encouraging the model to replicate the native style of the \Qibo library.
Our analysis suggests two key implications. First, the success of lightweight models ($<30$B parameters) confirms the viability of secure, cost-effective local deployment, achieving high accuracy without massive proprietary architectures. Second, this RAG-based evaluation functions as an effective documentation stress-test: by isolating modules with lower retrieval accuracy, maintainers can pinpoint gaps in the official documentation. This creates a virtuous cycle that improves library usability, a capability we further harness for automated documentation generation.

\subsection{Qualitative assessment of generated docstrings}
To validate the model across varied coding patterns, we applied the generation pipeline to a set of different \Qibo modules.
From this corpus, we select the \texttt{plot\_density\_hist} function (from \texttt{qibo.ui}) as a representative case study for detailed discussion, as illustrated in Figure \ref{fig:docstring_comparison}. The qualitative analysis of this sample highlights the potential of the RAG-assisted workflow:

\begin{figure*}[htbp]
    \centering
    \begin{minipage}[t]{0.48\textwidth}
        \centering
        \textbf{Human-Written Reference}
        \begin{lstlisting}[style=docstringStyle]
"""Plot the real and imaginary parts of the density matrix.

Given a :class:`qibo.models.circuit.Circuit`, plot the real and imaginary parts
of the final density matrix as separate 3D cityscape plots, side by side, and
with a gray ``z=0`` plane for the imaginary part.

Args:
    circuit (:class:`qibo.models.circuit.Circuit`): Circuit to visualize.
    title (str, optional): Title of the plot. Defaults to ``""``.
    [...] # (styling arguments omitted for brevity)
    backend (:class:`qibo.backends.abstract.Backend`, optional): backend
        to be used in the execution. If ``None``, it uses
        the current backend. Defaults to ``None``.
    n_most_relevant_components (int): Reduce ticks in x-axis. Default is None.

Returns:
    tuple: Respectively, the figure, and axes for the real and imaginary parts.
"""
        \end{lstlisting}
    \end{minipage}
    \hfill
    \begin{minipage}[t]{0.48\textwidth}
        \centering
        \textbf{QiboAgent-Generated (deepseek-r1:32b)}
        \begin{lstlisting}[style=docstringStyle]
"""
Visualizes the real and imaginary parts of a quantum state's density matrix.

Args:
    circuit (:class:`qibo.models.circuit.Circuit`): The quantum circuit whose state will be visualized.
    title (str, optional): Title for the plot. Default is "".
    [...]
    backend (:class:`qibo.backends.abstract.Backend`, optional): Backend to execute. If None, current backend is used.
    n_most_relevant_components (int, optional): Number of tick labels.

Returns:
    tuple: A tuple containing:
        - matplotlib.figure.Figure: The figure object.
        - matplotlib.axes.Axes: Axes for real part.
        - matplotlib.axes.Axes: Axes for imaginary part.

Examples:
    .. code-block:: python

        from qibo import Circuit, gates
        from qibo.ui import plot_density_hist

        circuit = Circuit(2)
        circuit.add(gates.H(0))
        circuit.add(gates.CNOT(0, 1))

        fig, ax_real, ax_imag = plot_density_hist(circuit)
"""
        \end{lstlisting}
    \end{minipage}
    \caption{\textbf{Side-by-side comparison of docstrings for \texttt{plot\_density\_hist}}. The model-generated version (right) includes precise return types and a functional code example, drawing on the RAG context to correctly instantiate the \texttt{Circuit} object.}
    \label{fig:docstring_comparison}
\end{figure*}

\begin{enumerate}
    \item \textbf{Contextual Accuracy via RAG:} A critical challenge is the correct cross-referencing of internal objects. In the primary case study, the model successfully linked the \texttt{Circuit} argument to \texttt{:class:`qibo.models.circuit.Circuit`}. While the retrieval mechanism minimizes hallucinations, handling complex inheritance structures remains a non-trivial task. However, compared to a standard LLM baseline, the Hybrid RAG pipeline enhances the generation of valid contextual references.

    \item \textbf{Content Enrichment:} The model-generated output complements the human baseline by systematically generating grounded and correct usage examples. Although the human-written description provides a more qualitative summary of the visual output ("cityscape plots"), the model's addition of a functional code snippet enhances the immediate usability of the documentation.
\end{enumerate}

It is crucial to frame these results within the intended operational scope of \QiboAgent. The system is not designed to replace human oversight but to serve as an intelligent assistant within the development lifecycle. The reliance on lightweight architectures, exemplified by the \texttt{deepseek-r1:32b} model used in this analysis, is key to this vision, as it ensures that the tool can be deployed locally within Continuous Integration (CI) pipelines with minimal resource overhead. This "human-in-the-loop" approach allows maintainers to receive and refine high-quality documentation proposals, selecting the best features from the AI generation while correcting subtle inaccuracies, thus significantly reducing the time burden of writing documentation from scratch.

\subsection{Resolution of Issue \#1710}

This section reports the results obtained by applying the agentic workflow of \QiboAgent to the specific task of autonomous pull request resolution. Utilizing the \texttt{gpt-oss:120b} model, we examine the system's performance in a real-world maintenance scenario through the detailed case study of Issue \#1710 \footnote{\url{https://github.com/qiboteam/qibo/issues/1710}}.

\begin{figure*}[htbp]
    \centering
    \begin{minipage}[t]{0.45\textwidth}
        \textbf{QiboAgent Framework output}
        \begin{lstlisting}[style=codeStyle]
def _apply_gate_unfold(self, gate, state, nqubits):
    state_tensor = self.np.reshape(state, (2,) * nqubits)
    targets = list(gate.qubits)
    remaining = [i for i in range(nqubits) if i not in targets]
    perm = targets + remaining
    state_perm = self.np.transpose(state_tensor, perm)
    dim_target = 2 ** len(targets)
    dim_rest = 2 ** (nqubits - len(targets))
    state_matrix = self.np.reshape(state_perm, (dim_target, dim_rest))
    matrix = gate.matrix(self)
    new_state_matrix = self.np.dot(matrix, state_matrix)
    new_state_perm = self.np.reshape(
        new_state_matrix,
        ([2] * len(targets)) + ([2] * (nqubits - len(targets)))
    )
    inv_perm = self.np.argsort(perm)
    new_state_tensor = self.np.transpose(new_state_perm, inv_perm)
    return self.np.reshape(new_state_tensor, (2 ** nqubits,))

def apply_gate(self, gate, state, nqubits):
    state = self.np.reshape(state, nqubits * (2,))
    matrix = gate.matrix(self)
    if gate.is_controlled_by:
        # ... (controlled gate logic omitted) ...
    else:
        return self._apply_gate_unfold(gate, state, nqubits)
    return self.np.reshape(state, (2**nqubits,))
        \end{lstlisting}
    \end{minipage}
    \hfill
    \begin{minipage}[t]{0.45\textwidth}
        \textbf{Human Verified Patch}
        \begin{lstlisting}[style=codeStyle]
@cache
def permutations(qubits: Tuple[int, ...], nqubits: int) -> Tuple[List[int], List[int]]:
    fwd_perm = list(qubits) + [q for q in range(nqubits) if q not in qubits]
    inv_perm = zip(list(range(nqubits)), fwd_perm)
    inv_perm, _ = list(zip(*sorted(inv_perm, key=lambda x: x[1])))
    return fwd_perm, inv_perm

def apply_gate(self, gate, state, nqubits):
    state = self.np.reshape(state, nqubits * (2,))
    matrix = gate.matrix(self)
    if gate.is_controlled_by:
        # ... (controlled gate logic omitted) ...
    else:
        shape = state.shape
        fwd_perm, inv_perm = einsum_utils.permutations(gate.qubits, nqubits)
        state = self.np.transpose(state, fwd_perm)
        state = self.np.reshape(state, (2 ** len(gate.qubits), -1))
        state = matrix @ state
        state = self.np.reshape(state, shape)
        state = self.np.transpose(state, inv_perm)
    return self.np.reshape(state, (2**nqubits,))
        \end{lstlisting}
    \end{minipage}
    \caption{\textbf{Side-by-side comparison of the AI-generated patch and the human-written patch.} The AI-generated code successfully implements the core logic of tensor unfolding and gate application, while the human patch incorporates additional optimizations such as caching permutation indices and utilizing the modern matrix multiplication operator.}
    \label{fig:patch_comparison}
\end{figure*}

A comparative analysis of the generated code, shown in Figure \ref{fig:patch_comparison}, confirms that the agent successfully derived the correct mathematical logic for tensor unfolding. As described in the previous section, the evaluation of the model's autonomous capabilities for this issue relied on two distinct runs. In the first run, where a conceptual code template was provided within the issue conversation, the agent seamlessly ingested and adapted the suggested structure into the codebase. More notably, in a second zero-shot run, where explicit code hints were deliberately withheld, the agent successfully abstracted the problem, independently formulating a mathematically and functionally correct solution from scratch.

However, when comparing the AI-generated code to the human-verified patch, qualitative differences emerge regarding specific architectural optimizations. The human implementation incorporates the \texttt{@cache} decorator via an external utility to compute the permutation indices only once, avoiding redundant calculations. In contrast, the agent's solution recalculates the permutation array and performs \texttt{np.argsort(perm)} upon every gate application, introducing computational overhead. Furthermore, while the agent elegantly encapsulates the logic into a dedicated private method (\texttt{\_apply\_gate\_unfold}), it relies on \texttt{np.dot} and explicit dimension calculations, whereas the human patch uses the modern \texttt{@} operator and implicit reshaping.

These findings highlight the intended role of \QiboAgent as a collaborative advanced assistant designed to alleviate developer workload. While the agent effectively manages routine maintenance tasks, such as the parameter setting request in Issue \#1699, it also proves highly capable in addressing architecturally complex challenges like Issue \#1710. The results are encouraging: the agent strictly follows developer instructions to produce a mathematically sound solution. The minor technical distinctions observed, such as the absence of caching, do not detract from the overall correctness of the patch. Instead, they demonstrate that the agent provides a solid, functionally complete implementation that can be directly approved or finalized with a reduced effort by the maintainers. Moreover, such refinements can be seamlessly addressed either by specifying performance constraints in the initial user prompt or through a brief iterative feedback loop following the patch proposal.

\subsection{Qibo-Core module generation}

We executed this generative task using the \texttt{gpt-oss:120b} model, explicitly selected for its superior instruction-following capabilities observed during the PR resolution experiments, while maintaining a computational footprint suitable for local deployment.

The experiment yielded a fully functional, albeit elementary, \texttt{Qibo-core} module \footnote{\url{https://github.com/qiboteam/qiboagent/tree/qibo_core_gen}}. The resulting artifact is not merely a collection of source files but a coherent project structure comprising the Rust source code, the Python bindings via \texttt{pyO3}, and the essential build configuration files (\texttt{Cargo.toml}, \texttt{pyproject.toml}). Furthermore, the system synthesized a suite of unit tests, which served as the objective function for the agentic loop.

\begin{figure*}[htbp]
    \centering
    \includegraphics[width=\textwidth]{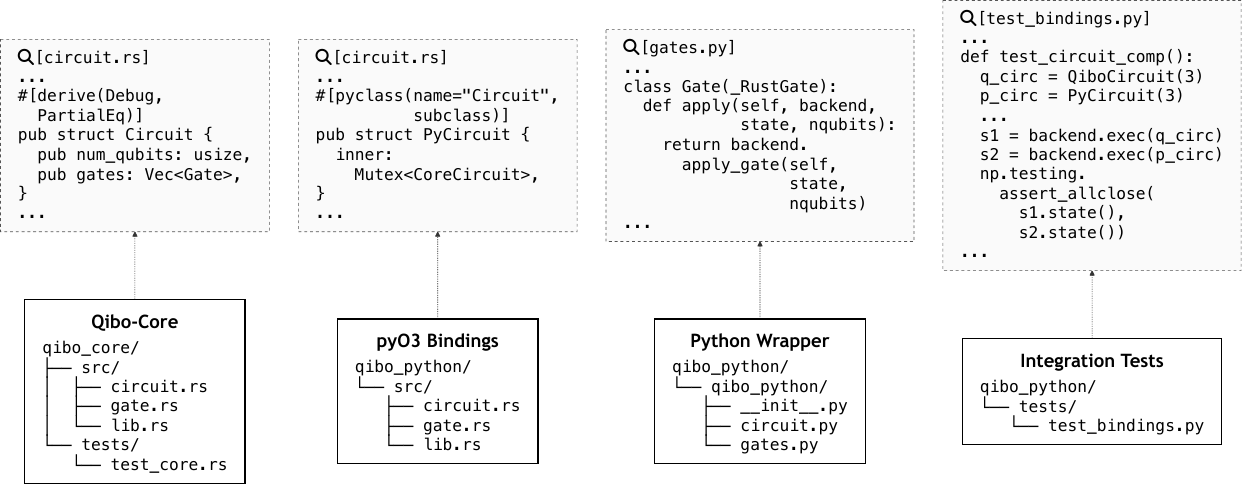}
    \caption{Overview of the generated \texttt{Qibo-core} project structure, accompanied by focused code snippets from different architectural layers. The integration test on the right highlights the verification process, ensuring full structural compatibility and execution equivalence with the existing \Qibo \NumPy backend.}
    \label{fig:core_result}
\end{figure*}

During the generation process, we observed that the Compiler-Driven Feedback Loop was the determining factor in the success of this task. In the initial iteration, the \textit{Rust Architect} and \textit{Binding Specialist} agents generated code that, while syntactically plausible, contained type-mismatch errors common in cross-language interfacing (\eg, incorrect ownership transfer between Python and Rust memory spaces).
Unlike standard generation pipelines where such errors would be terminal, the \textit{Debugger Agent} successfully parsed the compiler's error logs. It demonstrated the capacity to localize the fault and autonomously apply iterative patches until the compilation succeeded and the unit tests passed.

While the generated module is presented as a Proof of Concept and does not yet comprise all the functionalities supported by \Qibo today, it successfully executes fundamental quantum circuits. A key outcome of the experiment was the structural compatibility of the generated artifacts: the resulting \texttt{Circuit} and \texttt{Gate} Python objects function as drop-in replacements, fully compatible with existing \Qibo \NumPy backend. This achievement demonstrates the \QiboAgent architecture's potential to transcend the role of a passive assistant, acting instead as an active architect capable of handling complex, multi-step software modernization tasks in a code development scenario.

\section{Framework Availability}
\label{sec:availability}

The experimental evaluations in Section~\ref{sec:benchmark} highlight the advantages of Retrieval-Augmented Generation for static knowledge retrieval and Agentic workflows for autonomous coding tasks. Building on these results, we consolidated our experimental pipelines into a unified open-source framework: \QiboAgent. We offer two primary deployment options to ensure accessibility and flexibility, a standalone User Interface and a Model Context Protocol (MCP) integration.

The UI wraps the LangChain orchestration and the local Ollama inference engine, currently exposing two operational modes validated in our benchmarks:

\begin{itemize}
    \item \textbf{Interactive Q\&A (User Helper):} Based on the Question Benchmark (Sec.~\ref{sec:benchmark}), this chat interface connects to the Hybrid RAG pipeline. Users can query the \Qibo documentation in natural language, receiving and executing context-grounded code snippets.
    \item \textbf{Agentic Issue Resolution (Maintainer):} Following the pull request workflow tested in Sec.~\ref{sec:newfeaturepr}, this module allows users to input a GitHub issue reference. A monitoring dashboard lets users observe the agent's ``Reason and Act'' loop in real-time, tracking tool usage and the final code patch.
\end{itemize}

As an alternative to the UI, the MCP integration fits seamlessly into modern developer environments. By encapsulating our hybrid RAG pipeline within an MCP server, we decouple domain-specific knowledge retrieval from the user-facing application. Acting as an open standard that unifies communication between AI models and external tools, MCP allows developers to register the \QiboAgent backend with any compatible command line interface, giving users full flexibility in how they utilize the tools.

For a quick setup, we provide a fully configured environment in the project repository (\url{https://github.com/qiboteam/qiboagent}). We also include detailed system prompts and instruction sets for the agentic workflows to ensure full reproducibility.

\section{Closing Remarks}
\label{sec:conclusion}

In this work, we outlined a series of guidelines for integrating AI assistants into the lifecycle of quantum computing software, validated through the implementation of \QiboAgent. By moving beyond the paradigm of monolithic, general-purpose Large Language Models, we demonstrated that a domain-aware architecture, built upon the synergy of Retrieval-Augmented Generation and Agentic workflows, represents a robust and highly effective approach to the maintenance of complex scientific software.

Our experimental results validate two core hypotheses. First, regarding information access, our findings reinforce the necessity of a Hybrid Retrieval strategy for scientific domains. By fusing semantic understanding with strict keyword matching, the proposed pipeline effectively addresses the precision limitations of dense embeddings in this context, enabling accurate retrieval of rigid technical nomenclature and minimizing hallucination rates.
Second, regarding problem-solving, the successful resolution of real-world GitHub issues and the autonomous prototyping of the \texttt{Qibo-core} module show the robustness of the "Reason-Act" paradigm applied to scientific software maintenance.

We demonstrated that small-scale open-source models can deliver high-quality results without the need for massive computational resources. This makes local deployment a viable and efficient reality for research institutions that prioritize data privacy and independence from proprietary platforms.

Despite these promising results, our methodology presents certain limitations. The efficacy of the Retrieval-Augmented Generation pipeline heavily depends on the underlying knowledge base, where an unstructured or incomplete repository directly impairs the system's ability to synthesize accurate solutions. 

On the generative side, relying entirely on off-the-shelf lightweight models limits the system's ability to achieve the deep, intrinsic specialization that domain-specific fine-tuning could provide. Training a model directly on a specific codebase could certainly yield more highly optimized code for complex maintenance tasks. However, this is a highly costly and resource-intensive practice that demands a meticulously curated training pipeline to strictly prevent model overfitting. In addition, for an evolving repository, fine-tuning may ultimately prove ineffective: a model trained on current code states risks losing its generalized reasoning capabilities when confronted with future code changes, leading to a rapid degradation of performance. In contrast, the RAG-based approach provides a dynamic, up-to-date context that allows the model to adapt to changes in the codebase without requiring retraining.

Looking ahead, the successful integration of autonomous agents into the software maintenance lifecycle opens clear avenues for physical deployment. Given the potential demonstrated by these architectures, we intend to push \QiboAgent beyond software development tasks, applying its workflows to automate quantum hardware calibration via \Qibocal \cite{pasquale2024qibocalopensourceframeworkcalibration} and \Qibolab \cite{Efthymiou_2024}. This transition aims to elevate open-source AI assistants from repository maintainers to integral components of the full quantum computing stack.

\bibliographystyle{apsrev4-2}
\bibliography{references}

\end{document}